\documentclass[prl,twocolumn,superscriptaddress, showpacs]{revtex4}

\usepackage[dvips]{graphicx}
\usepackage{amssymb, amsmath, bbm}

\def\be{\begin{eqnarray}}
\def\ee{\end{eqnarray}}

\newcommand{\Eq}[1]{Eq.~(\ref{#1})}

\newcommand{\ra}{\rightarrow}

\pacs{74.72.-h, 75.10.JM }

\begin{document}
\title{Flux period, spin gap, and pairing in the one-dimensional $t-J-J'$-model}
\author{Alexander Seidel}
\affiliation{Department of Physics, University of California at Berkeley, Berkeley, CA 94720, USA}
\affiliation{Materials Sciences Division, Lawrence Berkeley National Laboratory }
\author{Dung-Hai Lee}
\affiliation{Department of Physics, University of California at Berkeley, Berkeley, CA 94720, USA}
\affiliation{Materials Sciences Division, Lawrence Berkeley National Laboratory }
\date{\today}
\begin{abstract}
Using the factorization of the wavefunction in the
$t$-$J$-$J'$-model at small exchange couplings, { we
demonstrate the connection between the existence of a spin gap and
an $hc/2e$ flux periodicity of the ground state energy. We conjecture that all spin-gapped SU(2)-invariant Luttinger liquids have $hc/2e$ flux periodicity,
and that this is connected to the fact that a gapped spin-$\frac
12$ chain always breaks translational symmetry by doubling the unit
cell.}
\end{abstract}
\maketitle

Soon after the discovery of high-$T_c$ { superconductivity,
Anderson proposed that the basic physics of the cuprates is that
of a doped two-dimensional Mott-insulator} \cite{PWA}. In
particular, the Cooper pairs of the superconducting { state}
are viewed as the ``liberated spin singlet pairs'' of the
insulating host material. { While this picture is very
attractive, it has been difficult to find an explicit model for
which the proclaimed behavior can be shown to occur
unequivocally.}\\
\indent { Searching for} models of Mott insulators that show
superconductivity upon doping has been the { motivation} for
{ many} studies of one-dimensional systems.\cite{OGSOAS, OGLURI,
IMADA, FABRIZIO, DARI, SEIDEL}
Thanks to { methods such as perturbative renormalization group
and bosonization}, considerable knowledge has been acquired on
the weak coupling phase diagram of both { strictly}
one-dimensional\cite{SOLYOM} and ladder systems\cite{linfisher}.
The drawback of the weak coupling approach is that it often is
only an instability analysis. The ultimate statement of the
quantum phase still { rests on certain assumptions} about the
``strong coupling fixed point'' of the renormalization group flow.
\\
\indent Most strong coupling models cannot be solved analytically.
A notable exception is the Luther-Emery action \cite{LE} which
describes an electronic liquid with a spin gap and dominant
singlet-superconducting (SS) correlations at large distances.
Another interesting analytic method for analyzing strong coupling
1D models was introduced by Ogata and Shiba \cite{OGSHI}, and
extended in Ref.\cite{OGLURI}. { This method is  designed} to
treat the large U Hubbard model (or the small $J$ $t$-$J$ model).
{ It is based on two facts:} i) in the limit of
$U\rightarrow\infty$ (or $J/t\ra 0$) the ground state of the
Hubbard ($t$-$J$ model) is infinitely degenerate, { and ii)
each of the degenerate states is described by a wavefunction
composed of a product of pure charge and spin components}
\cite{OGSHI}. For large but finite $U$ (small $J/t$) one can apply
degenerate perturbation theory to lift the degeneracy. After doing
so, { the ground state wavefunction remains factorized.
Moreover, the spin wavefunction is given by that of the Heisenberg
model on a ``squeezed lattice''(i.e. the lattice where the unoccupied sites are omitted)}.
\\
\indent In superconductivity the hallmark of electron pairing is
the $\Phi_0/2\equiv hc/2e$ flux period. In three dimensions, if
one plots the ground state energy $E(\Phi)$ of a solid
superconducting torus as a function of the Aharonov-Bohm (AB) flux
$\Phi$ { through} the hole, one finds a periodic function with
period $\Phi_0/2$. Moreover the energy barrier separating the
successive minima is extensive. { In two and one dimensions the
flux period is the same. However the energy barrier  becomes
intensive for two dimensions, and vanishes as the inverse
circumference for one dimension.}

In one dimension the spin and charge degrees of freedom decouple
in the low energy and long wavelength limit. { According to 
common wisdom, the presence of a spin gap implies pairing.} It is
thus natural to draw a connection between the { existence} of a
spin gap and a $\Phi_0/2$ flux period. However, since the vector
potential only enters in the charge action, it is not obvious how
the presence of a spin gap may affect the flux period. The purpose
of this paper is to clarify { this issue} in the context of a
strongly correlated 1D system.

{ In the following} we study the one-dimensional
$t$-$J$-$J'$-model, making use the degenerate perturbation
approach introduced in Ref.\cite{OGLURI}. The model is defined by
the Hamiltonian \vspace{-.2cm}
\begin{multline}\label{tJJ'}\vspace{-.4cm}
H =-t\sum_i {\cal P} (\,e^{\frac{2\pi i}{L}\frac{\Phi}{\Phi_0}}\,c^\dagger_{i,\sigma}c_{i+1,\sigma}+h.c.\,){\cal P} \\
+J\sum_i(S_i\cdot S_{i+1}-\frac{1}{4}n_in_{i+1})
  +J'\sum_i(S_i\cdot S_{i+2}-\frac{1}{4}n_in_{i+2}),
\end{multline}
describing N electrons on a ring of L sites in the presence of an
Aharonov-Bohm flux $\Phi$. Here, the projection operator $\cal P$
{ excludes states with doubly occupied sites}, and the $S_i$
are spin-$1/2$ operators. At $J=J'=0$ the model corresponds to the
$U=\infty$ Hubbard model. As pointed out in Ref.\cite{OGSHI}, in
this limit the eigenstates factorize into products of pure charge
and spin states. This property has been used extensively to study
the large (but finite) $U$ Hubbard model \cite{OGSHI, PARSOR,
OGSHI2}, which is related to the small $J$ $t$-$J$ model. Much
 less analytic work has been done { on the $t$-$J$-$J'$ model with a
finite $\alpha\equiv J'/J$, because the model is  no longer
integrable.} In this case, however, the degenerate perturbation
approach introduced in Ref.\cite{OGLURI} still allows one to
determine the ground state properties.


In the following we will use this method to study the
ground state energy of (\ref{tJJ'}) as a function of the AB-flux
$\Phi$. 
We begin by defining { the} $N$-particle wavefunction of the
system \be
\qquad\Psi(x_1,\sigma_1,...,x_N,\sigma_{N})=\left<0\left|c_{x_1\sigma_1}\dotsc
c_{x_N\sigma_N}\right|\Psi\right> \label{Psi}\ee on the domain
\\

$D:=\left\{(x_1\dotsc x_N)\in \mathbbm{Z}^N\right|
\left.x_1\!<\!x_2\!<\!\dotsc x_N\!<\! x_1+L\right\}.$ \\

\noindent { Here} $\left|\Psi\right>$ and $\left|0\right>$ are
the state of the system and the vacuum of the fermionic operators
$c_{x,\sigma}\equiv c_{x+L,\sigma}$, respectively. The fermion
antisymmetry and the periodic boundary condition imply
\vspace{-.3cm}
\begin{multline}\label{boundary}
\Psi\left(x_1,\sigma_1,...,x_N,\sigma_{N}\right)=\\
(-1)^{(N-1)}\,\Psi\left(x_2,\sigma_2,...,x_N,\sigma_{N}, (x_1+L),\sigma_1\right)\end{multline}
\indent At $J=J'=0$, { each eigen wavefunction of (\ref{tJJ'})
factorizes into a product of a charge and a spin wavefunction}
\cite{OGSHI}, 
\vspace{-.2cm}
\be
&&\Psi(x_1,\sigma_1,...,x_{N},\sigma_{N})=f(x_1\dotsc
x_N)\,g(\sigma_1\dotsc \sigma_N)\label{factor}
\vspace{-.2cm} 
\ee
\vspace{-.2cm}
where 
\vspace{-.2cm} 
\be
f(x_1\dotsc x_N)=\frac{1}{\sqrt{L^N}}\,\det\left[\,\exp(ik_i
x_j)\,\right] \label{slater}
\vspace{-.2cm} 
\ee
\noindent 
is a
{ Slater determinant constructed from $N$ plane waves.} It is
of central importance here to observe that for a finite ring with
periodic boundary conditions, the spin part and the charge part of
the wavefunction (\ref{factor}) are not completely independent.
{ Specifically} if we quantize the $N$ momenta in
(\ref{factor}) according to
\vspace{-.2cm}
\begin{equation}\label{kquant}
\begin{split}
&k_j=\frac{2\pi}{L}\, q_j+\frac{K}{L}\\
&\text{where } q_j\in \mathbbm{Z} \text{ and } K \in [0,2\pi),
\end{split}
\end{equation}
\noindent then the condition (\ref{boundary}) requires the spin
wavefunction to satisfy
\vspace{-.2cm}
\begin{equation}
g(\sigma_1\dotsc \sigma_N)=e^{iK}\,g(\sigma_2\dotsc
\sigma_N,\sigma_1),
\end{equation}
{\noindent which implies that $K$ is the ``spin momentum'' on the
squeezed lattice.} Thus by means of (\ref{kquant}), the momentum
$K$ of the spin wavefunction injects a twist into the charge wave
function. { For large but finite $U$ the exact ground state
wavefunction of the Hubbard model remains of the form given by
\Eq{factor}, and the same relation between charge twist and spin momentum is observed }\cite{OGSHI}.
Within the degenerate perturbation approach introduced in
Ref.\cite{OGLURI} the same { still} applies to the ground state of
(\ref{tJJ'}) for any value of $\alpha=J'/J$ in the limit of
vanishing exchange couplings. In this limit,{ 
 all solutions of
the form \Eq{factor}  are degenerate in the spin wavefunction, which is only required to have the spin momentum $K$ determined by the twist of the charge wavefunction.} 

\begin{figure}[h]
\begin{center}
\includegraphics[width=2in]{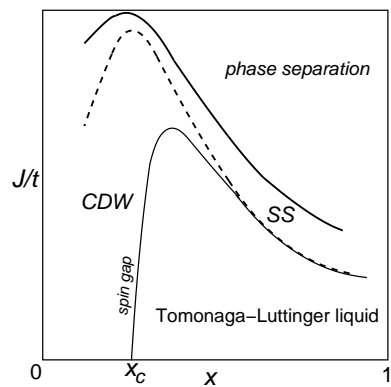}
\caption{\label{fig1} Sketch of the zero temperature phase diagram
of (\ref{tJJ'}) as obtained in \cite{OGLURI, NAKAMURA} for
$\alpha=\frac 12$. The spin-gapped region is divided by a
crossover (dashed) between regions of dominant singlet
superconducting (SS) and charge-density-wave (CDW) correlations.}
\vspace{-.7cm}
\end{center}
\end{figure}

To first order in the exchange couplings,
{ this} degeneracy is lifted by an effective Hamiltonian acting
in the ``squeezed'' space of $N$ spins \cite{OGLURI}:
\vspace{-.2cm}
\begin{eqnarray}
H_{eff}&=&\frac{L}{N}\left(J_{eff}\sum_{j=1}^{N}S_j\cdot S_{j+1}+ J'_{eff}\sum_{j=1}^{N}S_{j}\cdot S_{j+2}\right)\nonumber \\
J_{eff}&=&J\left<n_in_{i+1}\right>_{f}+J'\left<n_i(1-n_{i+1})n_{i+2}\right>_{f}\nonumber \\
J'_{eff}&=&J'\left<n_in_{i+1}n_{i+2}\right>_{f}
\label{Heff}
\end{eqnarray}
\noindent Here, $\left<\right>_{f}$ denotes a
spinless fermion expectation value with respect to the
 wavefunction $f$ displayed in (\ref{factor}).
\indent We now focus on the case of constant $\alpha$ where
$\alpha
>\alpha_c\approx .241$ \cite{JULHAL}. In this regime numerical and
analytical works suggest the phase diagram shown in Fig.
\ref{fig1}. At zero doping the spin chain corresponding to the
model (\ref{tJJ'}) at half filling ($N=L$) is gapped. The spin gap
will survive for a finite range of doping $x=1-N/L< x_c$ (Fig.
\ref{fig1}), and the effective spin Hamiltonian (\ref{Heff}) may
be used to calculate $x_c$ exactly in the limit $J/t\rightarrow
0$. This was 
first proposed by Ogata et al. \cite{OGLURI} and has
been confirmed numerically in Ref. \cite{NAKAMURA}. { Thus the
validity of degenerate perturbation theory is well established.
(We will give a detailed discussion of the involved subtleties
elsewhere \cite{prep} (see also \cite{SEIDEL}))}.\\
\indent To first order in $J/t$ and $J'/t$, the ground state
energy of \Eq{tJJ'} takes the form 
\be E_{tot}= E_c+E_s \label{E}
\vspace{-.2cm}
\ee  
\noindent where $E_c$ is the kinetic energy  associated with
the charge wavefunction $f_0$, and $E_s$ is the ground state
energy of (\ref{Heff}) { with the spin momentum $K$.} For any
given AB-flux, $f_0$
will be of the form given in (\ref{slater}),
where the $N$ consecutively occupied momenta $k_j$ are given by
(\ref{kquant}) with 
\vspace{-.2cm}
\be q_j=q_0+j-1,\quad j=1\dotsc N
\label{consec}
\ee
\begin{figure*}[tb]
\samepage
\begin{center}
\includegraphics[width=4.9in]{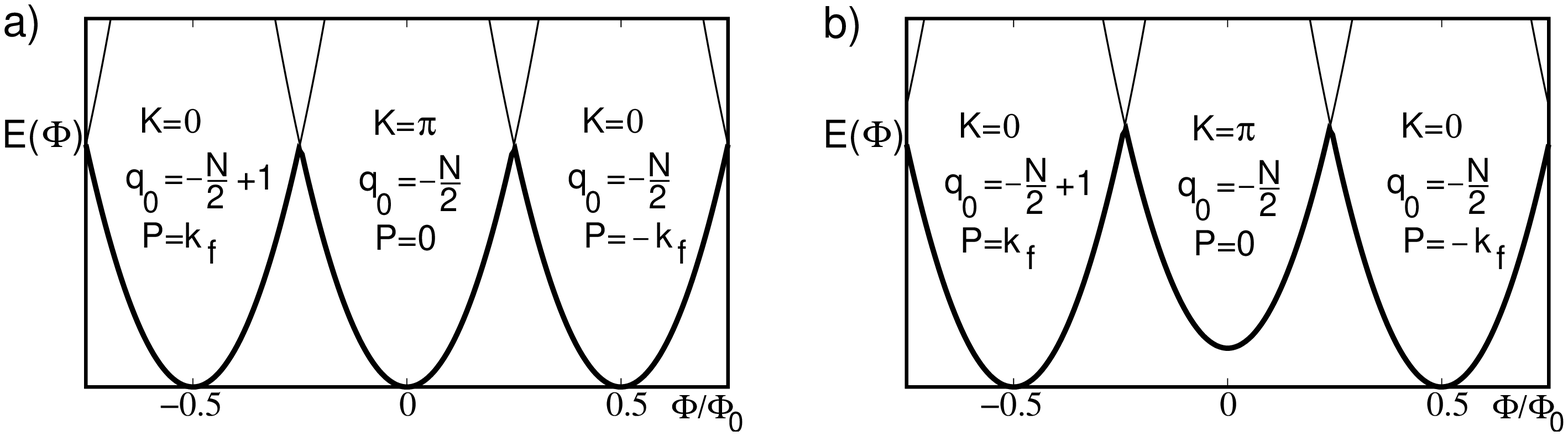}
\caption{\label{fig2} Ground state energy according to (\ref{E}), (\ref{E0})
for $x<x_c$ and $x>x_c$. The energy scale is $t/L$. a) $x<x_c$
(spin-gapped case). Periodic pattern of branches with $K=0$ and
$K=\pi$ separated by half a flux quantum. The thick line
represents the ground state energy at given $\Phi$. The ground state momentum
$P$ changes by $k_f$ between adjacent branches.
b)  $x>x_c$
(no spin gap). The $\frac 12 \Phi_0$-periodicity is destroyed by a
relative shift of order $J/L$ between $K=0$ and $K=\pi$ branches.}
\vspace{-.5cm}
\end{center}
\end{figure*}

\noindent and $q_0$ is an integer. The kinetic energy  is given by  
\begin{equation}\label{E0}
\begin{split}
&E_c(q_0,K,\Phi)=-2t\sum_j \,\cos\left(k_j+\frac{2\pi}{L}\frac{\Phi}{\Phi_0}\right)\\
&=-2t\,\frac{\sin(k_f)}{\sin\left(\frac{\pi}{L}\right)}\,\cos\left(k_f+\frac{2\pi}{L}q_0 +\frac{K-\pi}{L}+\frac{2\pi}{L}\frac{\Phi}{\Phi_0}\right)
\end{split}
\end{equation}
In \Eq{E0} $k_f$ is defined as $k_f\equiv\pi N/L$. Note that for
all charge wavefunctions characterized by \Eq{kquant},
(\ref{consec}) with different $q_0$ and $K$, the effective
couplings appearing in (\ref{Heff}) are the same.

\indent We will consider an even number of particles $N$ from now
on. In the spin-gapped regime $0<x<x_c$, { the effective
Hamiltonian (\ref{Heff}) has two lowest-energy states with spin
momenta $K=0$ and $K=\pi$. These two states are separated from
other spin states by an energy gap of order $J$ or $J'$. For a
finite ring, the energy difference between these two lowest spin
states vanishes exponentially with the circumference of the ring.
Thus in the thermodynamic limit these two states become
degenerate.}

 For fixed $\Phi$, we choose $K$ and $q_0$ so that the total
energy is minimized. This minimization can be achieved by first
minimizing $E_c$ by varying $q_0$ for fixed $K$, then minimizing
$E_s+E_c$ with respect to $K$. When the first minimization is
achieved the argument of the cosine is always of order $1/L$
regardless of the values of $K$. Hence the effect of varying $K$
in the second minimization can only result in $O(t/L)$ modulations
in $E_c$. { We consider the limit $t/L\ll J$ here. It then follows that 
the value of $K$ must be either $0$ or $\pi$.} Other choices of $K$ would
increase  $E_s$ by the spin gap of order $J$, which cannot be
compensated by the possible lowering of $E_c$.

After substituting the optimum value of $q_0$ for $K=0$ or $K=\pi$
 back into \Eq{E0} we obtain two branches of energy versus
$\Phi$ curves shown in Fig. \ref{fig2}a). The lower envelope of
these curves is the ground state energy as a function of $\Phi$
for small $J/t$ and $J'/t$. One observes that this function does
indeed show a period of $\Phi_0/2$, owing to the existence of two
different types of branches corresponding to $K=0$ and $K=\pi$,
respectively. This structure is resemblant of that proposed for
the dimer model \cite{kivelson,THOULESS}. It is also interesting
to note that $K=\pi$ is the analogy of the ``vison'' flux
\cite{z2} in 1D. 

\indent The situation is fundamentally different in the regime
$x>x_c$ (Fig. \ref{fig1}), where the spin gap vanishes. Here the
ground state of the effective spin Hamiltonian (\ref{Heff}) has
$K=0$ or $K=\pi$ depending on whether $N=4m$ or $N=4m+2$
(see e.g. \cite{TONEGAWA2}). For a finite chain of length $N=O(L)$ the first
spin excited states at $K=\pi$ or $K=0$ (i. e. whose momenta
differ by $\pi$ from that of the respective ground state) have
 excitation energies of order $J/L$. This energy gives rise to the relative shift between the $K=0$ and $K=\pi$
 branches shown in Fig. \ref{fig2}b) for $N=4m$. The resulting lower envelope is illustrated for a small
 but finite $J/t$, where the flux period is now $\Phi_0$. For
$N=4m+2$ the shift between the $K=0$ and $K=\pi$ branches is
opposite in sign. This behavior is well demonstrated in the large
repulsive $U$ Hubbard model \cite{FEKULA}, where no spin gap is
present.

It is sometimes felt 
that the existence of metastable minima of $E(\Phi)$ at $\Phi_0/2$
intervals is the sign of a pairing tendency, even though $\Phi_0/2$
is strictly not the flux period. { Fig.\ref{fig2}b) presents a
clear counter example of this type of reasoning. Indeed according
to Fig. \ref{fig2}b) this can happen} in a regime of the phase
diagram where the ground state is neither spin-gapped, nor
features a dominance of superconducting pairing correlations.

\indent On the other hand, for $x<x_c$, there is a spin gap but no
charge gap, and the ground state energy is a periodic function of
$\Phi$ with period $\Phi_0/2$. It has thus all the characteristics
of a superconductor. However for $J/t\ll 1$ the {
superconducting correlations (SS) are weaker than
the charge density wave (CDW) correlations} (Fig. \ref{fig1}). In
this regime we can think of the system as being close to a
superconductor-(Cooper pair) insulator transition due to strong
quantum fluctuation of the phase of the superconducting order
parameter. Here, a weak external perturbation such as disorder can
easily localize the Cooper pairs and drive the system insulating.
As the system crosses the crossover line in (Fig. \ref{fig1}), the
phase fluctuations become much less { severe} so that the SS
correlations become dominant over the CDW correlations. We expect the 
appearance of a $\Phi_0/2$ flux period to hold in the entire spin
gapped regime.
Indeed, more detailed considerations show that the arguments given
here are not limited to first order perturbation theory \cite{prep}.
In particular, we find that second order contributions to the
ground state energy may be incorporated into the effective spin
Hamiltonian. The latter will then also depend on the
twist of the zeroth order wave function and on flux, yet 
$H_{eff}(K,\Phi)=H_{eff}(K-\pi,\Phi+\Phi_0/2)$ continues to hold.
Furthermore, we have shown that all the results presented here
also follow from a weak coupling/bosonization procedure \cite{bosonize}.

\indent The analysis presented here can be applied to a wide class
of models of the form \Eq{tJJ'} where the second line is replaced
by a more general spin-chain type of Hamiltonian. Most features of
the phase diagram shown in Fig. \ref{fig1} will likely survive as
long as the spin chain at half filling is gapped. In particular,
the spin gap will survive for a range of doping, and phase
separation will occur at sufficiently large values of $J/t$, where
$J$ is an appropriate energy scale for the spin couplings. As the
phase separation line is approached, the charge compressibility
diverges, and Luttinger liquid physics then implies a regime of
dominant SS correlations. { Our analysis on flux period will
then carry over to this more generic case}, {\em provided that}
the gapped spin state at $x=0$ also breaks translational symmetry
by doubling of the unit cell, analogous to the dimerization that
occurs in the { $J$-$J'$} model at half filling. Such an example
is given by the $t$-$J_z$ model studied in Ref. \cite{FERRARI}.
Although in Ref. \cite{FERRARI}, the possibility of $\Phi_0/n$ 
flux periods $(n\geq 2)$ has been postulated for models of the 
type considered here, only the case $n=2$ has been found for the
$t$-$J_z$ model. This follows easily along the line of arguments
given here, and we believe that only $n=1$ and $n=2$ are found
in generic models.

\indent 
If, on the other hand, a gapped spin-$\frac 12$ chain
exists that does not break translational symmetry, it appears that a
doped model with a spin gap could be constructed which {\em does
not} feature $\Phi_0/2$ flux quantization as displayed in Fig.
\ref{fig2}a). However, such a state would violate the Lieb-Schultz-Mattis
theorem \cite{LSM}. 
For SU(2)-invariant spin-$\frac 12$ { chains} in one
dimension, we are aware of only one way to create a spin gap, i.e.
breaking the translational symmetry by doubling the unit cell. Hence there seems to be an intimate relation between this fact and the possible universality of the $\Phi_0/2$ flux period which we postulate below.\\ 
\indent We note that a $\Phi_0/2$ 
flux period associated with a spin gap has also been observed in numerical 
studies of a two-leg ladder \cite{HAYWARD}. This suggests that our 
main conclusion may be generalized beyond the purely one-dimensional case. 
However, a
two-leg ladder has an even number of sites per unit cell. Here the undoped
system may have a spin gap due to the formation of singlet pairs located on
the rungs, which does not require symmetry breaking. These singlet pairs become
mobile upon doping, and the above notion of symmetry breaking in some internal spin space is not required to explain the $\Phi_0/2$ period. 

\indent To conclude, { we have demonstrated the relation
between a $\Phi_0/2$-periodicity in the ground state energy and
the existence of a spin gap in the small exchange limit of the
$t$-$J$-$J'$ model.} Based on these findings, we conjecture that the observed $\Phi_0/2$ flux period is a universal property of spin-gapped SU(2) invariant one-dimensional systems of spin-$\frac 12$ particles with gapless charge degrees of freedom. In particular, the value of the charge Luttinger parameter is not a determining factor of
the flux periodicity, as our result did not require the predominance of singlet superconducting correlations. Our findings further suggest an intimate relation between the proposed universality of the $\Phi_0/2$ flux period and the fact that all gapped SU(2)-invariant spin-$\frac 12$ chains feature broken translational symmetry with a doubling of the unit cell.
\begin{acknowledgments}
\vspace{-.0cm}
This work has been supported by DOE grant DE-AC03-76SF00098.
\end{acknowledgments}
\vspace{-.7cm}

\begin{thebibliography}{26}
\expandafter\ifx\csname natexlab\endcsname\relax\def\natexlab#1{#1}\fi
\expandafter\ifx\csname bibnamefont\endcsname\relax
  \def\bibnamefont#1{#1}\fi
\expandafter\ifx\csname bibfnamefont\endcsname\relax
  \def\bibfnamefont#1{#1}\fi
\expandafter\ifx\csname citenamefont\endcsname\relax
  \def\citenamefont#1{#1}\fi
\expandafter\ifx\csname url\endcsname\relax
  \def\url#1{\texttt{#1}}\fi
\expandafter\ifx\csname urlprefix\endcsname\relax\def\urlprefix{URL }\fi
\providecommand{\bibinfo}[2]{#2}
\providecommand{\eprint}[2][]{\url{#2}}

\bibitem[{\citenamefont{Anderson}(1987)}]{PWA}
\bibinfo{author}{\bibfnamefont{P.~W.} \bibnamefont{Anderson}},
  \bibinfo{journal}{Science 235, 1196}  (\bibinfo{year}{1987}).

\bibitem[{\citenamefont{Ogata et~al.}(1991{\natexlab{a}})\citenamefont{Ogata,
  Luchini, Sorella, and Assaad}}]{OGSOAS}
\bibinfo{author}{\bibfnamefont{M.}~\bibnamefont{Ogata}},
  \bibinfo{author}{\bibfnamefont{M.~U.} \bibnamefont{Luchini}},
  \bibinfo{author}{\bibfnamefont{S.}~\bibnamefont{Sorella}}, \bibnamefont{and}
  \bibinfo{author}{\bibfnamefont{F.~F.} \bibnamefont{Assaad}},
  \bibinfo{journal}{Phys. Rev. Lett. 66, 2388}
  (\bibinfo{year}{1991}{\natexlab{a}}).

\bibitem[{\citenamefont{Ogata et~al.}(1991{\natexlab{b}})\citenamefont{Ogata,
  Luchini, and Rice}}]{OGLURI}
\bibinfo{author}{\bibfnamefont{M.}~\bibnamefont{Ogata}},
  \bibinfo{author}{\bibfnamefont{M.~U.} \bibnamefont{Luchini}},
  \bibnamefont{and} \bibinfo{author}{\bibfnamefont{T.~M.} \bibnamefont{Rice}},
  \bibinfo{journal}{Phys. Rev. B 44, 12083}
  (\bibinfo{year}{1991}{\natexlab{b}}).

\bibitem[{\citenamefont{Imada}(1993)}]{IMADA}
\bibinfo{author}{\bibfnamefont{M.}~\bibnamefont{Imada}},
  \bibinfo{journal}{Phys. Rev. B 48, 550}  (\bibinfo{year}{1993}).

\bibitem[{\citenamefont{Fabrizio}(1996)}]{FABRIZIO}
\bibinfo{author}{\bibfnamefont{M.}~\bibnamefont{Fabrizio}},
  \bibinfo{journal}{Phys. Rev. B 54, 10054}  (\bibinfo{year}{1996}).

\bibitem[{\citenamefont{Dagotto and Rice}(1996)}]{DARI}
\bibinfo{author}{\bibfnamefont{E.}~\bibnamefont{Dagotto}} \bibnamefont{and}
  \bibinfo{author}{\bibfnamefont{T.~M.} \bibnamefont{Rice}},
  \bibinfo{journal}{Science 271, 618}  (\bibinfo{year}{1996}).

\bibitem[{\citenamefont{Seidel and Lee}(2004{\natexlab{a}})}]{SEIDEL}
\bibinfo{author}{\bibfnamefont{A.}~\bibnamefont{Seidel}} \bibnamefont{and}
  \bibinfo{author}{\bibfnamefont{P.~A.} \bibnamefont{Lee}},
  \bibinfo{journal}{Phys. Rev. B 69, 094419}
  (\bibinfo{year}{2004}{\natexlab{a}}).

\bibitem[{\citenamefont{{S\'{o}lyom}}(1979)}]{SOLYOM}
\bibinfo{author}{\bibfnamefont{J.}~\bibnamefont{{S\'{o}lyom}}},
  \bibinfo{journal}{Adv. Phys. 28, 201}  (\bibinfo{year}{1979}).

\bibitem[{\citenamefont{Lin et~al.}(1997)\citenamefont{Lin, Balents, and
  Fisher}}]{linfisher}
\bibinfo{author}{\bibfnamefont{H.-H.} \bibnamefont{Lin}},
  \bibinfo{author}{\bibfnamefont{L.}~\bibnamefont{Balents}}, \bibnamefont{and}
  \bibinfo{author}{\bibfnamefont{M.~P.~A.} \bibnamefont{Fisher}},
  \bibinfo{journal}{Phys. Rev. B 56, 6569}  (\bibinfo{year}{1997}).

\bibitem[{\citenamefont{Luther and Emery}(1974)}]{LE}
\bibinfo{author}{\bibfnamefont{A.}~\bibnamefont{Luther}} \bibnamefont{and}
  \bibinfo{author}{\bibfnamefont{V.~J.} \bibnamefont{Emery}},
  \bibinfo{journal}{Phys. Rev. Lett. 33, 589}  (\bibinfo{year}{1974}).

\bibitem[{\citenamefont{Ogata and Shiba}(1990)}]{OGSHI}
\bibinfo{author}{\bibfnamefont{M.}~\bibnamefont{Ogata}} \bibnamefont{and}
  \bibinfo{author}{\bibfnamefont{H.}~\bibnamefont{Shiba}},
  \bibinfo{journal}{Phys. Rev. B 41, 2326}  (\bibinfo{year}{1990}).

\bibitem[{\citenamefont{Parola and Sorella}(1990)}]{PARSOR}
\bibinfo{author}{\bibfnamefont{A.}~\bibnamefont{Parola}} \bibnamefont{and}
  \bibinfo{author}{\bibfnamefont{S.}~\bibnamefont{Sorella}},
  \bibinfo{journal}{Phys.Rev. Lett. 64, 1831}  (\bibinfo{year}{1990}).

\bibitem[{\citenamefont{Ogata et~al.}(1991{\natexlab{c}})\citenamefont{Ogata,
  Sugiyama, and Shiba}}]{OGSHI2}
\bibinfo{author}{\bibfnamefont{M.}~\bibnamefont{Ogata}},
  \bibinfo{author}{\bibfnamefont{T.}~\bibnamefont{Sugiyama}}, \bibnamefont{and}
  \bibinfo{author}{\bibfnamefont{H.}~\bibnamefont{Shiba}},
  \bibinfo{journal}{Phys. Rev. B 43, 8401}
  (\bibinfo{year}{1991}{\natexlab{c}}).

\bibitem[{\citenamefont{Jullien and Haldane}(1983)}]{JULHAL}
\bibinfo{author}{\bibfnamefont{R.}~\bibnamefont{Jullien}} \bibnamefont{and}
  \bibinfo{author}{\bibfnamefont{F.~D.~M.} \bibnamefont{Haldane}},
  \bibinfo{journal}{Bull. Am. Phys. Soc. 28, 344}  (\bibinfo{year}{1983}).

\bibitem[{\citenamefont{Nakamura}(1998)}]{NAKAMURA}
\bibinfo{author}{\bibfnamefont{M.}~\bibnamefont{Nakamura}},
  \bibinfo{journal}{J. Phys. Soc. Jpn. 67, 717}  (\bibinfo{year}{1998}).

\bibitem[{\citenamefont{Seidel and Lee}()}]{prep}
\bibinfo{author}{\bibfnamefont{A.}~\bibnamefont{Seidel}} \bibnamefont{and}
  \bibinfo{author}{\bibfnamefont{D.-H.} \bibnamefont{Lee}},
  \bibinfo{howpublished}{in preparation}.

\bibitem[{\citenamefont{Kivelson et~al.}(1987)\citenamefont{Kivelson, Rokhsar,
  and Sethna}}]{kivelson}
\bibinfo{author}{\bibfnamefont{S.~A.} \bibnamefont{Kivelson}},
  \bibinfo{author}{\bibfnamefont{D.~S.} \bibnamefont{Rokhsar}},
  \bibnamefont{and} \bibinfo{author}{\bibfnamefont{J.~P.}
  \bibnamefont{Sethna}}, \bibinfo{journal}{Phys. Rev. B 35, 8865}
  (\bibinfo{year}{1987}).

\bibitem[{\citenamefont{Thouless}(1987)}]{THOULESS}
\bibinfo{author}{\bibfnamefont{D.~J.} \bibnamefont{Thouless}},
  \bibinfo{journal}{Phys. Rev. B 36, 7187}  (\bibinfo{year}{1987}).

\bibitem[{\citenamefont{Senthil and Fisher}(2000)}]{z2}
\bibinfo{author}{\bibfnamefont{T.}~\bibnamefont{Senthil}} \bibnamefont{and}
  \bibinfo{author}{\bibfnamefont{M.~P.~A.} \bibnamefont{Fisher}},
  \bibinfo{journal}{Phys. Rev. B 62, 7850}  (\bibinfo{year}{2000}).

\bibitem[{\citenamefont{Tonegawa and Harada}(1987)}]{TONEGAWA2}
\bibinfo{author}{\bibfnamefont{T.}~\bibnamefont{Tonegawa}} \bibnamefont{and}
  \bibinfo{author}{\bibfnamefont{I.}~\bibnamefont{Harada}},
  \bibinfo{journal}{J. Phys. Soc. Jpn. 56, 2153}  (\bibinfo{year}{1987}).

\bibitem[{\citenamefont{Ferretti et~al.}(1992)\citenamefont{Ferretti, Kulik, and
  Lami}}]{FEKULA}
\bibinfo{author}{\bibfnamefont{A.}~\bibnamefont{Ferretti}},
  \bibinfo{author}{\bibfnamefont{I.~O.} \bibnamefont{Kulik}}, \bibnamefont{and}
  \bibinfo{author}{\bibfnamefont{A.}~\bibnamefont{Lami}},
  \bibinfo{journal}{Phys. Rev. B 45, 5486}  (\bibinfo{year}{1992}).

\bibitem[{\citenamefont{Seidel and Lee}(2004{\natexlab{b}})}]{bosonize}
\bibinfo{author}{\bibfnamefont{A.}~\bibnamefont{Seidel}} \bibnamefont{and}
  \bibinfo{author}{\bibfnamefont{D.-H.} \bibnamefont{Lee}},
  \bibinfo{journal}{cond-mat/0402663}  (\bibinfo{year}{2004}{\natexlab{b}}),
  \bibinfo{note}{to be published}.


\bibitem[{\citenamefont{Ferrari and Chiappe}(1996)}]{FERRARI}
\bibinfo{author}{\bibfnamefont{V.}~\bibnamefont{Ferrari}} \bibnamefont{and}
  \bibinfo{author}{\bibfnamefont{G.}~\bibnamefont{Chiappe}},
  \bibinfo{journal}{J. Phys.: Condens. Matter 8, 8583}  (\bibinfo{year}{1996}).

\bibitem[{\citenamefont{Lieb et~al.}(1961)\citenamefont{Lieb, Schultz, and
  Mattis}}]{LSM}
\bibinfo{author}{\bibfnamefont{E.~H.} \bibnamefont{Lieb}},
  \bibinfo{author}{\bibfnamefont{T.~D.} \bibnamefont{Schultz}},
  \bibnamefont{and} \bibinfo{author}{\bibfnamefont{D.~C.}
  \bibnamefont{Mattis}}, \bibinfo{journal}{Ann. Phys. 16, 407}
  (\bibinfo{year}{1961}).

\bibitem[{\citenamefont{Hayward et~al.}(1995)\citenamefont{Hayward, Poilblanc,
  Noack, Scalapino, and Hanke}}]{HAYWARD}
\bibinfo{author}{\bibfnamefont{C.~A.} \bibnamefont{Hayward}},
  \bibinfo{author}{\bibfnamefont{D.}~\bibnamefont{Poilblanc}},
  \bibinfo{author}{\bibfnamefont{R.~M.} \bibnamefont{Noack}},
  \bibinfo{author}{\bibfnamefont{D.~J.} \bibnamefont{Scalapino}},
  \bibnamefont{and} \bibinfo{author}{\bibfnamefont{W.}~\bibnamefont{Hanke}},
  \bibinfo{journal}{Phys. Rev. Lett. 75, 926}  (\bibinfo{year}{1995}).

\end{thebibliography}

\end{document}